\documentstyle[12pt]{article}
\date{}
\title{The cluster-core model for halo-structure of \\
light nuclei at the drip lines.}
\author{Raj K. Gupta$^{a,b}$, Sushil Kumar$^{a}$, M. Balasubramaniam$^{a,b}$\\
G. M\"unzenberg$^c$ and Werner Scheid$^b$,\\
{\it $^a$ Physics Department, Panjab University, Chandigarh-160014,}\\
{\it India.}\\
{\it $^b$ Institut f\"ur Theoretische Physik, Justus-Liebig-Universit\"at,}\\
{\it Heinrich-Buff-Ring 16, D-35392, Giessen, Germany.}\\
{\it $^c$ Gesellschaft f\"ur Schwerionenforschung mbH, Planckstrasse 1,} \\
{\it D-64291 Darmstadt, Germany.}\\
}

\begin{document}
\maketitle
\def\be{\begin{equation}}
\def\ee{\end{equation}}
\def\ba{\begin{array}}
\def\ea{\end{array}}
\def\bea{\begin{eqnarray}}
\def\eea{\end{eqnarray}}
\begin{abstract}
Nuclei at both the neutron- and proton-drip lines are studied. In
the cluster-core model, the halo-structure of all the observed and
proposed cases of neutron- or proton-halos is investigated in
terms of simple potential energy surfaces calculated as the sum of
binding energies, Coulomb repulsion, nuclear proximity attraction
and the centrifugal potential for all the possible cluster+core
configurations of a nucleus. The clusters of neutrons and protons
are taken to be unbound, with additional Coulomb energy added for
proton-clusters. The model predictions agree with the available
experimental studies but show some differences with the nucleon
separation energy hypothesis, particularly for proton-halo nuclei.
Of particular interest are the halo-structures of $^{11}N$ and
$^{20}Mg$. The calculated potential energy surfaces are also
useful to identify the new magic numbers and molecular structures
in exotic nuclei. In particular, N=6 is a possible new magic number
for very neutron-deficient nuclei, but Z=N=2 and Z=8 seem to
remain magic even for such nuclei, near the drip line.

\end{abstract}

\baselineskip 18pt
\newpage

\section{Introduction}
The halo-structure refers to highly neutron-rich or proton-rich
{\it light} nuclei that lie, respectively, near the neutron- or
proton-drip line and hence are totally "unstable" systems. A
number of such nuclei have now become available, both as the
primary and secondary beams with various low, intermediate, and
high energies, called the radioactive nuclear beams (RNBs). The
structure properties of some of these nuclei are studied via the
projectile fragmentation process in reactions using the
intermediate and high energy RNBs. These structures are found to
be different from the earlier known structures of nuclei at or
near the $\beta$-stability line, and are referred to as halo or
thick neutron (proton) skin structures,
depending on the measurements made (see below).
So far only 6-7 cases of
neutron-halo and 3-4 cases of proton-halo nuclei are identified
experimentally, but many more are proposed as the likely
candidates. These are listed in Tables 1 and 2, respectively, for
neutron- and proton-halo nuclei.

A halo or thick-skin nucleus is identified by its measured weak
binding or small neutron (proton) separation energy, extended
nucleon density distributions $\rho(r)$ ($\rho _n$ for n-halo and
$\rho _p$ for p-halo, a model dependent property) or large
root-mean-square radii (not following the $R=R_0A^{1\over 3}$ law;
for p-rich nuclei the enlargement of matter radii is observed to
be not as much as for n-rich nuclei) and narrow momentum
distributions of the emitted neutron(s) or proton(s) (observed to
be more so for n-halo than for p-halo).
The terminology thick neutron- or proton-skin is used when the
nucleon density distribution is measured, representing the
thickness of the extended neutron or proton surface. The other
properties characterize these nuclei as halo nuclei which is
more of an artistic view and is useful for developing a
few-body theory (see below). Here, we note that the last two
properties (extension in space and narrow momentum distribution
of, say, neutrons)
are related, since as
an obvious consequence of the Heisenberg's uncertainity principle
$(\Delta x  \Delta p \sim \hbar),$ a large extension in space must
restrict the momentum to a narrow distribution.
Apparently, the term halo or thick-skin serves for book keeping only.
For the 1n and 2n
separation energies defined, in terms of the binding energy
B(Z,N), as
\be S_{1n} = M(Z,N-1) + M_n - M(Z,N) = B(Z,N) - B(Z,N-1)
\label{eq:1}
\ee
and
\be
S_{2n} = B(Z,N) - B(Z,N-2),
\label{eq:2}
\ee
a nucleus is 1n- or 2n-halo nucleus if $S_{1n}$
or $S_{2n}$ is the lowest and is less than 1 MeV (to be compared with
6$-$8 MeV for stable nuclei). The same is found to be nearly true for
p-halo nuclei, i.e.
\be
S_{1p} = M(Z-1,N) + M_p - M(Z,N) = B(Z,N) - B(Z-1,N)
\label{eq:3}
\ee
or
\be
S_{2p} = B(Z,N) - B(Z-2,N),
\label{eq:4}
\ee
is the lowest and less than 1 MeV, though the observed
number of cases for p-halo are so far only a few. The 2n-halo
nucleus is found to possess an additional characteristic property,
namely: its isotope with one less neutron (for example, $^{10}Li$
for $^{11}Li$) is also unbound \cite{young94}. Based on these
characteristic properties, the experimentally observed 1n-halo
nuclei are $^{11}Be$ and $^{19}C$
\cite{anne93,kelley95,marques96,bazin98} and the 2n-halo nuclei
are $^{6,8}He$, $^{11}Li$, $^{14}Be$ and $^{17}B$
\cite{tani85,tani92,saint89,orr95,alkhazov97}. Since the existence
of a bound di-neutron in the ground-state is not yet observed
\cite{ieki96,sauvan01} (a free di-neutron is also unbound by about
70 KeV), apparently, we are dealing here with two-body (n+core)
and three-body (n+n+core) structures, respectively, for the 1n-
and 2n-halo nuclei. Table 1 shows that many other cases have been
investigated, both theoretically and experimentally
\cite{tani96,villari91,hansen95,bergmann01,navin00}, and in some
calculations \cite{ren96} even a 3n-halo structure is predicted
(for $^{26}F$) which means a four-body system. Thus, it is evident
that {\it a priori} information on the halo-structure of a nucleus
is of vital importance for the theoretical treatment of these
weakly bound nuclei as two-, three- or more-body problems
\cite{thompson99}. Similar remarks apply to proton-halo nuclei,
where 1p-halo structures are established for $^8B$
\cite{schwab95,guimaraes00} and $^{17}F$ \cite{ozawa94} and the
2p-halo for only $^{17}Ne$ \cite{ozawa94}. For $^{11}N$, the
experiments seem to indicate a 1p-halo structure
\cite{oliveira00}, but in the following we find that this nucleus
could be another case of a 4-body system (a 3p-halo structure).
Table 2 shows many other experimental or theoretical possible
cases \cite{brown96,ren96a,ren96b}, which includes even a
possibility of the tetra-proton cluster surrounding the $^{16}O$
core proposed for $^{20}Mg$ \cite{chulkov96}.

The direct experimental information on the halo-structures of nuclei
come from the Coulomb breakup studies or the nuclear dissociation
mechanism (Coulomb dissociation of projectile in the presence of heavy
targets) \cite{marques96,bazin98,orr95,labiche01}.
The breakup process could either be a "direct
process" or occur via a low energy (soft) dipole mode (a giant resonance
phenomenon) between the core and skin- or halo-nucleons
\cite{takigawa91,dasso92}, in addition to the normal dipole resonance where
the neutron and proton cores are separated. However, the Coulomb dissociation
experiments \cite{sackett93,nakamura94} seem to rule out the possibility of a
soft dipole resonance and suggest that the breakup of halo nuclei is
a "direct breakup" process. But, in a very recent charge exchange reaction
$^6Li(^7Li,^7Be)^6He$ \cite{nakayama00}, a resonance is observed
whose excitation energy, width and cross section seem consistent with those
expected for a soft dipole resonance in $^6He$. On the other hand, the measured
cross sections of an earlier charge changing reaction for
$^{8,9,11}Li$ on C target strongly suggest that $^{11}Li$ is a three-body
system with $^9Li$ as core and the two neutrons orbiting around it
\cite{blank92}. These results further strengthen the picture where the halo
neutrons in, for example, $^{11}Li$ are taken to exist as unbound, rather
than bound in a di-neutron. In other words, instead of a resonant state, an
excited state of nonzero width is formed which has important consequences
for the structure studies of neutron-rich nuclei
\cite{csoto94,zhukov94,esbensen93}.

In this paper, we dwell on the question of how to characterize,
{\it a priori}, a nucleus as a halo-nucleus i.e. find the
halo-structure of a nucleus. We discuss here a simple theoretical
method of potential energy surfaces (PES), the cluster-core model
(CCM), given recently by some of us \cite{gupta95,gupta2k} for
neutron-rich light nuclei. This was refereed to as (neutron)
cluster-core model. We find that the method works equally well for
proton-rich light nuclei, provided the Coulomb energy of protons
in proton-clusters is added. This is now given the name (nucleon)
cluster-core model; in short the CCM. Also, the new cases of
possible neutron-halo are studied. It may be mentioned that this
is the only known theoretical method for identifying the
halo-structure of a n- or p-rich light nucleus, other than the
hypothesis of (neutron or proton) separation energy described in
the second paragraph above.

The (nucleon) cluster-core model (CCM) is described in section 2, with the results
of our calculations for both the n- and p-halo nuclei presented in section 3.
The possibility of identifying the newly arisen magic numbers in exotic nuclei,
within the same cluster-core model, is also discussed in this section.
A summary of our results is given in section 4.

\section{The (nucleon) cluster-core model (CCM)}
As already stated above, the CCM was first given \cite{gupta95,gupta2k} for
the neutron-clusters, and is now extended to proton-clusters also.
Note that the cluster is now a general term, used not only for
$\alpha$-particle and heavier nuclei but also for one or two protons, since
the proton-radioactivity is also understood on the basis of the models used
for $\alpha$-decay and exotic cluster radioactivity (see e.g. \cite{buck92}).
Here, we extend the same concept to neutron-clusters also, which seems to
work equally well.
In the CCM, we calculate the potential energy surface (PES) of a nucleus
for its all possible cluster-core ($A_2,A_1$) configurations and look for a
neutron(s)-cluster (or proton(s)-cluster) + core configuration with a minimum
potential energy, which in the language of the cluster-decay model (e.g. the
Preformed Cluster Model of Gupta et. al. \cite{gupta88,malik89,gupta91,kumar97})
means a configuration formed with the largest quantum mechanical probability.
Such a picture is consistent with the model of Hansen and Jonson \cite{hansen87}
where our cluster-core configuration could be understood as their quasi-deutron
picture of $^{11}Li$, approximated as $^{9}Li$ core plus a di-neutron
consisting of two {\it unbound} neutrons. Alternatively, our cluster-core
configurations could be looked upon as few-body configurations of the
type used in few-body theoretical calculations \cite{thompson99}. The calculated
PES in CCM could also be used for the search of optimum molecular structures,
reached in experiments via the break-up reactions \cite{leask01}.

The  potential energy for a cluster-core configuration ($A_2,A_1$) of a
nucleus A (=$A_1+A_2$; $A_1$ and $A_2$ being, respectively, the core and
cluster mass numbers) is defined simply as the sum of the two binding energies,
the Coulomb repulsion, the additional attraction due to nuclear proximity and
the rotational energy due to angular momentum:
\be
V(A_1, A_2, R, \ell ) = -\sum_{i=1}^{2} B(A_{i}, Z_{i}) + \frac{Z_{1} Z_{2} e^{2}}{R}
+ V_{P} +V_{\ell}.
\label{eq:5}
\ee
Here, $B(A_i,Z_i)$ are the experimental binding energies or mass excesses
$\Delta m_A = M_A - A$ in energy units \cite{audi95} (the two quantities
differ through a constant since $M_A(Z,N)=ZM_p+NM_n-B(Z,N)$), $V_P$ the
nuclear proximity interaction energy \cite{blocki77} and $V_{\ell}$ the
centrifugal potential \cite{saroha85} that pulls the fragments apart.
The deformation effects are neglected here in both the Coulomb and proxomity
energies, for reasons of simplicity only. These effects are shown to be
important for a complete quantitative cluster decay model calculation,
specifically the cluster preformation probability \cite{kumar97}.
Also, it may
be pointed out that here we are using experimental binding energies and
hence the microscopic shell effects are there in Eq. (\ref{eq:5}). However,
it will be of interest to see the role of macroscopic liquid drop energy
alone for the halo-structure. Like the $\alpha$-nucleus structure in
light-heavy nuclei is due to the Wigner term \cite{sharma2k,gupta2k1}, the
halo-structure in light nuclei at the drip-lines could be due to the surface
energy term in the liquid drop formula.

The charges $Z_i$ in Eq. (\ref{eq:5}) are fixed by minimizing the sum of the
two binding energies in $Z_1$ (or $Z_2$). Also, the shape of the potential
$V(A_1,A_2, R, \ell )$ is known to be nearly independent of the choice of the value of
R.
This is shown explicitly in many of our earlier publications (see e.g. in
\cite{sandu76} for early works or in \cite{gupta99} for a review).
Therefore, in view of our dealing here with light nuclei,
we consider only touching configurations, i.e., $R=R_1+R_2=R_t$. In general, we define
$R_i=R_0A_i^{1/3}$, where $R_0$ has a constant value (=1.15 fm). This means
that $R_i$ depend only on $A_i$, i.e., $R_i(A_i)$. If we allow surface effects
and define the S\"ussman central radius $C_i=R_i-(1/{R_i})$, we still get
$C_i = C_i(A_i)$. However, for halo nuclei $R_0$ changes considerably from
nucleus to nucleus and from one isotope to another isotope. Therefore, we
consider $R_0(Z_i,A_i)$ and take (or interpolate) its values for $Z\le 10$
from the experimental \cite{saint89,ozawa94} or theoretical \cite{grue96}
data and for $Z>10$ from the theoretical estimates of \cite{brown96}.
This means, we include the surface effects not only through S\"ussman central
radii but also via the observed isotopic variations of the radii in halo
nuclei. For the neutron radius, we approximate it to be the same as the
root-mean-square (rms) radius of proton, namely 0.8 fm \cite{ozawa94}.

The binding energy for a cluster with $x$ neutrons ($x\ge 1$) is
taken to be $x$ times that of the one-neutron binding energy (the
mass excess $\Delta m_n = 8.0713$ MeV), i.e.
\be
B(A_2=xn) = x\Delta m_n.  \qquad\qquad {\hbox {(for neutron-clusters)}}
\label{eq:6}
\ee
For proton-clusters, we define the same as
\be
B(A_2=xp) = x\Delta m_p - a_c A_2^{5\over 3}, \qquad {\hbox {(for proton-clusters)}}
\label{eq:7}
\ee
with $\Delta m_p = 7.2880$ MeV, the one-proton mass excess (equivalent of the
one-proton binding energy), and $a_c$=0.7053 MeV \cite{myers66}. The
additional term in (\ref{eq:7}) is the disruptive Coulomb energy
($=- a_c({Z_2^2/A_2^{1\over 3}})$) between the $x$ protons (here
$x=A_2=Z_2$). The above definitions for the binding energies of n-
or p-clusters mean that the nucleons in these clusters are taken
to be unbound, just as is the case in the model of Hansen and
Jonson \cite{hansen87}, the few-body theories \cite{thompson99} or
as is suggested by the recent experiments \cite{ieki96,sauvan01}.

\section{Calculations and results}
We have calculated the potential energy surfaces $V(A_2)$ for all the
neutron- and proton-rich nuclei listed in Tables 1 and 2, by using the
experimental binding energies from the 1995 Tables of Audi and Wapstra
\cite{audi95} and $R_0=R_0(Z_i,A_i)$ in $R_i=R_0 A_i^{1/3}$. Several of the
n-halo nuclei listed in Table 1 were already studied in our earlier works
\cite{gupta95,gupta2k}, and we discuss here the additional new cases
proposed experimentally or theoretically. We find that all the nuclei
considered here in Tables 1 and 2, except $^{11}N$, are stable ($Q<0$)
against all possible cluster-core ($A_2$,$A_1$) configurations in the ground
state. $^{11}N$ is unstable against all proton-cluster decays (such clusters
are 1p-, 2p- and 3p-clusters). Also, the
cluster-core configurations corresponding to the minima in these potential
energy surfaces are the most probable cluster-core configurations formed for
the cluster-decay process \cite{gupta88,gupta99a}, i.e., they occur with relatively
larger preformation probabilities, compared to their neighbors. Of these
cluster-core configurations, we are interested here only in the one(s) where
a cluster of neutrons (or protons) is involved (the other configurations are
also of interest, but more for seeking new magic numbers and molecular
structures, etc.). Such a neutron(s) or proton(s) cluster will behave like a
n- or p-halo since this is most loosely bound to the core. We find that the
nuclei considered here are all either 1nucleon- or 2nucleon-halo nuclei, and
we discuss them in the following separately for the two cases of n- and
p-halo structures.

For the angular momentum part of the potential, we have considered both the
cases of $\ell =0$ and $\ell \ne 0$ ($\ell =1,2,3$, chosen arbitrarily).
In general, both the positions and depths of potential energy minima in
$V(A_2)$ are {\it nearly} independent of the contribution of the
$\ell$-dependent term in the potential \cite{saroha85}, but we find that in
some cases, like for $^{11}N$, $^{12}N$ and $^{24}O$, there is a shift
of the minimum to another nucleon-cluster. This point needs further
investigation and perhaps refers to the mixed angular momentum and parity
states for the ground state configuration \cite{simon99,dattap00}. For this
reason, in the following, we discuss only the $\ell =0$ configurations.

\subsection{Neutron-halo nuclei}
Table 1 lists all the cases of neutron-halo studied here, as well as their
one-neutron and two-neutron separation energies, $S_{1n}$ and $S_{2n}$, the
configurations with respect to the minimum in the PES calculated with the CCM and
the N:Z ratios of the resulting core nuclei. The nuclei studied earlier in
\cite{gupta2k} are also included here in this table. We find that in agreement
with our earlier results of \cite{gupta2k}, the cores could be classified
mostly as the ones with N=2Z and $2Z\pm 2$ nuclei. This means that
nuclei with  $2Z\pm 2$ are as stable as the $N=2Z$ nuclei. The cores with
$2Z-3$ and $2Z-4$ are also predicted by the CCM, but we notice in Table 1 that
according to the neutron separation energy hypothesis, there is a question
mark with the halo-structure of at least some of these nuclei (see footnotes
in Table 1); e.g. for $^{22}O$ and $^{24}F$ both the neutron separation
energies $S_{1n}$ and $S_{2n}$ are much larger than $\sim$1 MeV. The halo
nuclei with larger separation energies ($>>$1 MeV) are perhaps good cases of
collective soft dipole excitations, though the same is not realized
experimentally as yet \cite{leistenschnei01}.

Figure 1 shows the results of our calculation for the four new cases, two each
of 1n- and 2n-halo structures. We notice in Fig. 1 (upper frames) and Table 1
(upper part; for the corresponding PES of the earlier studied cases, see Fig.
1 in \cite{gupta2k}) that for all the 1n-halo nuclei, the deepest minimum
clearly occurs at the 1n+core configuration. The minimum in the PES means
the most probable n-cluster+core configuration for the nucleus. Apparently,
these nuclei can be considered as two-body (n+core) systems \cite{thompson99}.
On the other hand, Table 1 shows that the neutron separation energy in all
the 1n-halo nuclei is though minimum for $S_{1n}$, but is $>>$1 MeV for some
cases ($^{22}O$, $^{23}O$, $^{24}O$ and $^{24}F$). These are, however,
exotic nuclei, lying far from the $\beta$-stability line, and hence should be
good candidates of halo structure
\cite{villari91,ren96,leistenschnei01,ozawa00}.

Figure 1 (lower frames) and Table 1 (lower part) show the results of our
calculation for 2n-halo nuclei. In these nuclei, the neutron(s)-cluster
minimum in the PES for each case is deepest at the 2n+core configuration and,
except for $^{12}Be$ and $^{27}F$, the $S_{2n}$ is also the lowest. For both
the $^{12}Be$ and $^{27}F$, the lowest neutron separation energy is for
one-neutron separation $S_{1n}$, whereas the PES for these nuclei show the
2n+core configuration lying clearly lowest (for $^{12}Be$, also both $S_{1n}$
and $S_{2n}>>1$ MeV). On the other hand, $^{29}F$ is clearly a 2n-halo nucleus
from the point of view of the separation energy ($S_{2n}$ is lowest, compared to
$S_{1n}$ and $S_{3n}$) but in the PES, the minima at 3n+ and 4n+core are almost
as deep as for the 2n+core configuration (see Fig. 2 \cite{gupta2k}). Hence,
$^{29}F$ seems to be an isolated n-halo nucleus where neutron-clusters which
are more complex than a di-neutron could exist, as was also expected by Migdal
\cite{migdal73}. However, we have listed in Table 1 the $^{12}Be$, $^{27}F$
and $^{29}F$ nuclei as 2n-halo nuclei from the point of view of the PES.
Experimentally also, the $^{12}Be$ nucleus is expected \cite{navin00} to have
a 2n-halo structure, identical to that of the classical 2n-halo nucleus
$^{11}Li$. Another exception to the separation energy hypothesis is $^8He$
where both $S_{1n}$ and $S_{2n}>>1$ MeV. This nucleus is now a well studied
halo-nucleus \cite{alkhazov97} and is predicted to be a 2n-halo structure in
the CCM. The other experimentally observed $^{6}He$, $^{14}Be$ and $^{17}B$
are also the well studied structures of 2n-halo nuclei
\cite{tani85,saint89,orr95}. Three-body (n+n+core) calculations have also
been made for some of these nuclei \cite{thompson99,garrido99,arai99}.

\subsection{Proton-halo nuclei}
Table 2 lists all the cases of 1p-, 2p- and possibly 3p-halo nuclei studied
here. Figures 2 and 3 present the results of our calculation, respectively,
for 1p- and 2p-halo nuclei. $^{11}N$, the only case of predicted 3p-halo, is
also included amongst the 1p-halo nuclei, since both experiments
\cite{oliveira00} and proton-separation energy hypothesis characterize it as
a 1p-halo structure.

The proton-halo structures are so far observed only in a few cases
(4 cases: $^8B$, $^{11}N$, $^{17}F$ and $^{17}Ne$), because of the
suppression of the p-halo cluster minimum in PES  by
the Coulomb forces; the Coulomb repulsion is zero for neutrons and
hence the n-halo configurations lie lower. Once again we discuss
here only the case of $\ell =0$, since the $\ell$-contribution to
both the positions and relative depths of potential energy minima
is found small, except for the two $^{11,12}N$ nuclei (see Fig.
2).

Table 2 and Fig. 2 (excluding the $^{11}N$ nucleus) show that for all the
1p-halo nuclei, the results of our CCM calculation (the PES minima) agree
completely with those of the $S_{1p}$ hypothesis. This includes the two
observed cases of $^{8}B$ and $^{17}F$ \cite{schwab95,ozawa94} and the three
experimentally or theoretically proposed cases of $^{12}N$, $^{26}P$ and
$^{27}P$ \cite{tani96,brown96,ren96a}. The theoretical predictions are based
on the shell model \cite{brown96} and RMF \cite{ren96a} calculations.
For the 2p-halo nuclei, however, our theoretical method of PES disagrees with
the proton separation energy hypothesis in more than one way. For $^9C$,
$^{18}Ne$ and $^{27}S$, whereas the proton-separation energy in Table 2 is the
lowest for $S_{1p}$, the PES in Fig. 3 show clear minima at 2p+core
configurations. The shell model \cite{brown96} and RMF \cite{ren96b}
calculations also predict both the $^{18}Ne$ and $^{27}S$ to be the 2p-halo
nuclei. Also, for $^{18}Ne$ (and so also for $^{20}Mg$) both $S_{1p}$ and
$S_{2p}$ are $>>1$ MeV. The $^{20}Mg$ nucleus, which is a 2p-halo structure
on the basis of the PES (Fig. 3), is a case of special interest. This is
predicted as 4 protons circulating the $^{16}O$ core on the basis of Coulomb
displacement energies \cite{chulkov96}. The CCM calculations show that the
4p+core configuration is also one of the possible configurations in $V(\eta_Z)$
(see inset, Fig. 3, $^{20}Mg$) but is not the most preferred one (does not lie
at the minimum either in $V(\eta_Z)$ or in $V(A_2)$). Also, the RMF calculations for $^{20}Mg$
\cite{ren96b} support the 2p-halo configuration, as is predicted by the CCM.

The $^{11}N$ nucleus is listed in Table 2 as a 1p- or 3p-halo structure. This
is a completely unstable nucleus against proton-decays, since $Q>0$ for its
1p-, 2p- and 3p-decays. The $S_{1p}$ is negative and lowest which means that
it is a 1p-halo nucleus according to the proton-separation energy hypothesis.
The experimental observation of its ground state and excitation spectra also
give it a 1p-halo configuration \cite{oliveira00}. However, the PES in Fig. 3
present it as a clear 3p-halo structure. Apparently, further
measurements and calculations are essential, since as per present calculations
this nucleus seems to be the first candidate of a rather more complex p-halo
structure.

Finally, the core configurations for p-halo nuclei consist of Z=N or Z=N+1,
which means that Z=N+1 core is as stable as the Z=N one. Also, cores with
Z=N+2 and N+3 are predicted but these are the cases where the halo structure is
still to be confirmed experimentally.

\subsection{New magic numbers in exotic nuclei}
We consider here only the case of $^{20}Mg$, as an illustrative example, and
show that N=6 arises as a magic number for the very neutron-deficient, exotic
nuclei only.

We know that in a potential energy surface, like the ones shown in Figs. 1-3,
the minima arise only due to the shell effects of either one or both the
fragments \cite{gupta99,gupta99a}. This means that at the potential energy
minima, at least one of the contributing nucleus must be a magic or nearly
magic nucleus. In Fig. 3, the calculated PES for $^{20}Mg$ gives four minima,
where both the participating nuclei are shown. We notice that for the very
neutron-deficient nuclei near the drip-line, like $^{16}Ne$ and $^{14}O$, N=6
could be the magic number but for the neutron-deficient nuclei, like $^{18}Ne$,
lying not too far from the $\beta$-stability line, N=8 is still the magic number.
The N=2 and Z=2 magicities seem to be kept as such for both the classes (near and
far away from $\beta$-stability line) of nuclei. Also, Z=8 seems to remain magic for
both kinds of nuclei, since the proton numbers combining with all the
N=6 (or its neighboring) nuclei are with Z=8 or its neighbors with Z=$8\pm 2$. This
is an interesting result and needs further investigations on the basis of the CCM
as well as other methods \cite{ozawa00,otsuka01,raj01}.

\section{Summary}

Summarizing, we have extended the cluster-core model (CCM), given
by some of us \cite{gupta95,gupta2k} earlier for (unbound)
neutron-clusters, to include also the (unbound) proton-clusters.
In other words, the cluster-core model could now be used to find,
a priori, the halo-structure of any light nucleus approaching the
$\beta^+$- or $\beta^-$-instability, i.e., the light neutron- and
proton-rich nuclei near the neutron- and proton-drip lines. This
knowledge is important for building any few-body theory of these
weakly bound systems. The CCM is applied to all known cases,
experimentally established and/or theoretically proposed as the
likely candidates, of neutron- and proton-halos. For proton-halo
nuclei, an additional Coulomb energy is found essential for the,
otherwise unbound, proton-clusters. In comparison to the only
other known hypothesis of neutron (or proton) separation energy,
the present model works well without any exception.
A theoretical method is in any case preferred over an (empirical)
hypothesis and this is the merit of the CCM presented here.
The
predictions of the two methods (CCM and nucleon separation energy
hypothesis) differ in many cases and this is more so for
proton-halo nuclei. One such interesting case is of 3p-halo
structure predicted by the CCM for a completely proton-unbound
$^{11}N$ nucleus. $^{20}Mg$ is perhaps another interesting nucleus
where both $S_{1p}$ and $S_{2p}$ are $>>$1 MeV and the CCM predicts
it to be a 2p-halo structure in contradiction to its expected
$4p+^{16}O$ configuration.

The calculated potential energy surfaces (PES) in CCM are also useful for
searching the possible new magic numbers and optimum molecular structures
for exotic nuclei. Our first analysis of the PES for $^{20}Mg$ nucleus
allows us to identify N=6 as the possible magic number for very neutron-deficient
nuclei near the drip line, though N=8 is still shown to be a magic number
for neutron-deficient nuclei not too far from the line of $\beta$-stability. Also,
Z=N=2 and Z=8 are found to remain magic even for very neutron-deficient nuclei.\\

\par\noindent
{\bf Acknowledgments:}\\
RKG, MB and WS are thankful to Volkswagen-Stiftung, Germany, for the support
of this research work under a Collaborative Reserach Project between the
Panjab University and Giessen University. RKG and MB are also thankful to the
Council of Scientific and Industrial Research (CSIR), New Delhi, for the
partial support of this research work. SK is thankful to Department of Atomic
Energy (DAE), Govt. of India for a Junior Research Fellowship.

\newpage
\begin{table}[t]
{\bf Table 1:}
{The calculated halo-characteristics of light neutron-rich nuclei. The neutron
separation energies are calculated by using the binding energy tables of
\cite{audi95} and the configurations resulting from the potential
energy surfaces (PES) are with respect to the $\ell=0$ case.
For observed cases, see text.

}
\begin{center}
\begin{tabular}{|l|c|c|c|c|l|} \hline
 Structure &Nucleus &$S_{1n}$      &$S_{2n}$  &PES Minimum    &Core  \\
           &        &(KeV)      &(KeV)     &(Cluster+core Config.)&(N:Z)  \\ \hline
 1n-halo  &$^{11}Be$ &504       &7317      &$1n+^{10}Be$   &N=2Z-2 \\
          &$^{14}B$  &970       &5848      &$1n+^{13}B$    &N=2Z-2 \\
          &$^{15}C$  &1218      &9394      &$1n+^{14}C$    &N=2Z-4 \\
          &$^{17}C$  &729       &4979      &$1n+^{16}C$    &N=2Z-2 \\
          &$^{19}C$  &162       &4346      &$1n+^{18}C$    &N=2Z   \\
          &$^{22}N$  &1222      &5828      &$1n+^{21}N$    &N=2Z   \\
          &$^{22}O$  &6848$^*$  &10655     &$1n+^{21}O$    &N=2Z-3 \\
          &$^{23}O$  &2740$^*$  &9588      &$1n+^{22}O$    &N=2Z-2  \\
          &$^{24}O$  &3713$^*$  &6453      &$1n+^{23}O$    &N=2Z-1  \\
          &$^{24}F$  &3857$^*$  &11392     &$1n+^{23}F$    &N=2Z-4  \\
          &$^{26}F$  &1050      &5399      &$1n+^{25}F$    &N=2Z-2  \\
          &$^{29}Ne$ &1330      &5216      &$1n+^{28}Ne$   &N=2Z-2  \\ \hline
 2n-halo  &$^{6}He$  &1864      &974       &$2n+^{4}He$    &N=2Z-2  \\
          &$^{8}He$  &2583$^*$  &2138      &$2n+^{6}He$    &N=2Z    \\
          &$^{11}Li$ &326       &301       &$2n+^{9}Li$    &N=2Z    \\
          &$^{12}Be$ &3169$^{*,a}$&3673    &$2n+^{10}Be$   &N=2Z-2  \\
          &$^{14}Be$ &1847      &1336      &$2n+^{12}Be$   &N=2Z    \\
          &$^{17}B$  &1437      &1393      &$2n+^{15}B$    &N=2Z    \\
          &$^{19}B$  &1029      &495       &$2n+^{17}B$    &N=2Z+2  \\
          &$^{22}C$  &1448      &1120      &$2n+^{20}C$    &N=2Z+2  \\
          &$^{27}F$  &1309$^a$  &2359      &$2n+^{25}F$    &N=2Z-2  \\
          &$^{29}F^b$&1002      &897       &$2n+^{27}F$    &N=2Z    \\
\hline
\end{tabular}
\end{center}
$^*$ Note that the neutron separation $S_{1n}$ (or $S_{2n}$) is much larger than $\sim$1 MeV.\\
$^a$ The $S_{1n}$-value is lower than the $S_{2n}$-value, whereas in the PES
the 2n minimum is clearly the lowest.\\
$^b$ The $S_{2n}$-value is lower and the minimum in PES
also lies at 2n+core configuration, though in the PES,
for both $\ell=0$ and $\ell\ne 0$ cases, the 2n, 3n and 4n minima are very
close to each other. However, the $S_{3n}$-value (2206 KeV) is higher than
both the $S_{1n}$ and $S_{2n}$ energies.
\end{table}

\newpage
\begin{table}[t]
{\bf Table 2:}
{Same as for Table 1, but for proton-rich nuclei.
}
\begin{center}
\begin{tabular}{|l|c|c|c|c|l|} \hline
 Structure &Nucleus &$S_{1p}$      &$S_{2p}$  &PES Minimum    &Core  \\
           &        &(KeV)      &(KeV)     &(Cluster+Core Config.)&(N:Z)    \\ \hline
 1p-halo  &$^{8}B$   &137       &5743      &$1p+^{7}Be$    &Z=N+1  \\
          &$^{12}N$  &601       &9290      &$1p+^{11}C$    &Z=N+1 \\
          &$^{17}F$  &600       &12727     &$1p+^{16}O$    &Z=N   \\
          &$^{26}P$  &141       &3549      &$1p+^{25}Si$   &Z=N+3  \\
          &$^{27}P$  &897       &6415      &$1p+^{26}Si$   &Z=N+2  \\ \hline
 2p-halo  &$^{9}C$   &1296$^a$  &1433      &$2p+^{7}Be$    &Z=N+1  \\
          &$^{17}Ne$ &1484      &948       &$2p+^{15}O$    &Z=N+1  \\
          &$^{18}Ne$ &3932$^b$  &4534      &$2p+^{16}O$    &Z=N+1  \\
          &$^{20}Mg$ &2647$^b$  &2314      &$2p+^{18}Ne$   &Z=N+2  \\
          &$^{27}S$  &755$^a$   &896       &$2p+^{25}Si$   &Z=N+3  \\ \hline
1p/3p-halo&$^{11}N$  &-1973$^*$ &2033      &$3p+^{8}Be$    &Z=N \\
          \hline
\end{tabular}
\end{center}
$^*$ Note that here $S_{1p}$ is negative, which means a completely unstable
system against 1p-decay ($Q_{1p}$-value=1.974 MeV). However, the PES prefers
a 3p-halo and it is also unstable against 3p-decay ($Q_{3p}$-value=2.556
MeV). Also, it is unstable against 2p-decay ($Q_{2p}$-value=0.208 MeV).
Furthermore, $S_{3p}$=1.848 MeV, which means $S_{3p}<S_{2p}$.\\
$^a$ Though $S_{1p}<S_{2p}$, the PES suggests this nucleus as 2p-halo.\\
$^b$ Both $S_{1p}$ and $S_{2p}$ are $>>$1 MeV.\\

\end{table}

\newpage
\par\noindent
{\bf Figure Captions}\\

\par\noindent
Fig. 1 {The potential energy V as a function of the light cluster mass
$A_2$ for 1n- and 2n-halo nuclei, showing the deepest minimum at 1n- (or 2n-)
cluster + core configurations. For $R_0(Z_i,A_i)$, see text.
}\\

\par\noindent
Fig. 2 {The same as for Fig. 1, but for 1p- and 3p-halo nuclei, showing
the deepest minimum at 1p (or 3p)-cluster+core configurations. The
3p-cluster+core configuration is predicted only for $^{11}N$ nucleus.
}\\

\par\noindent
Fig. 3 {The same as for Fig. 2, but for 2p-halo nuclei, showing
the deepest minimum at 2p-cluster + core configurations.
}\\

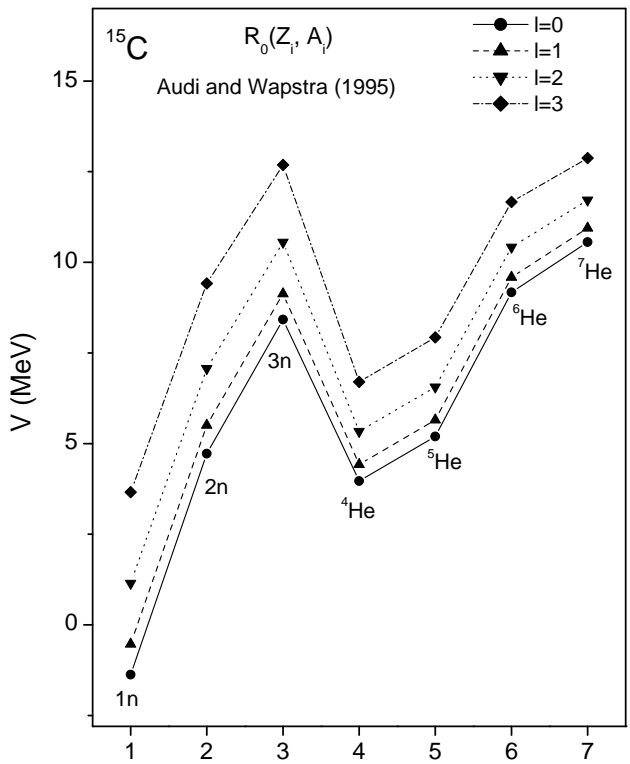
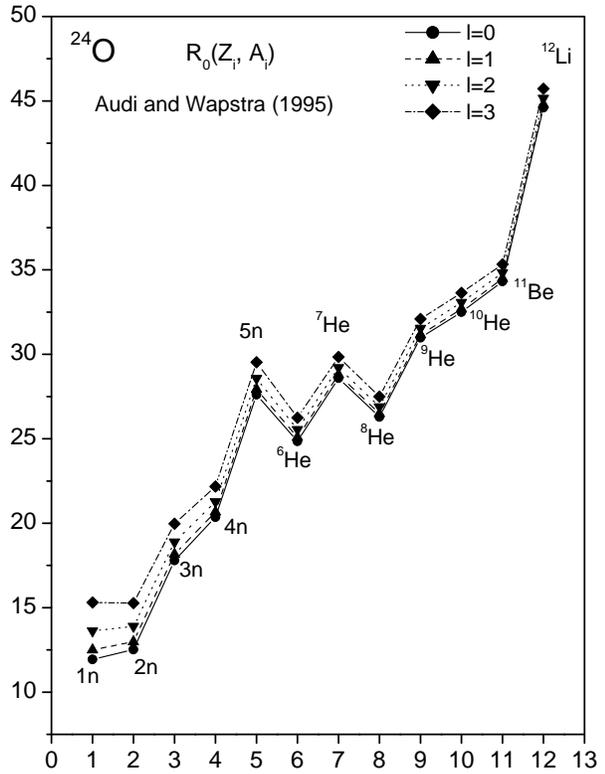
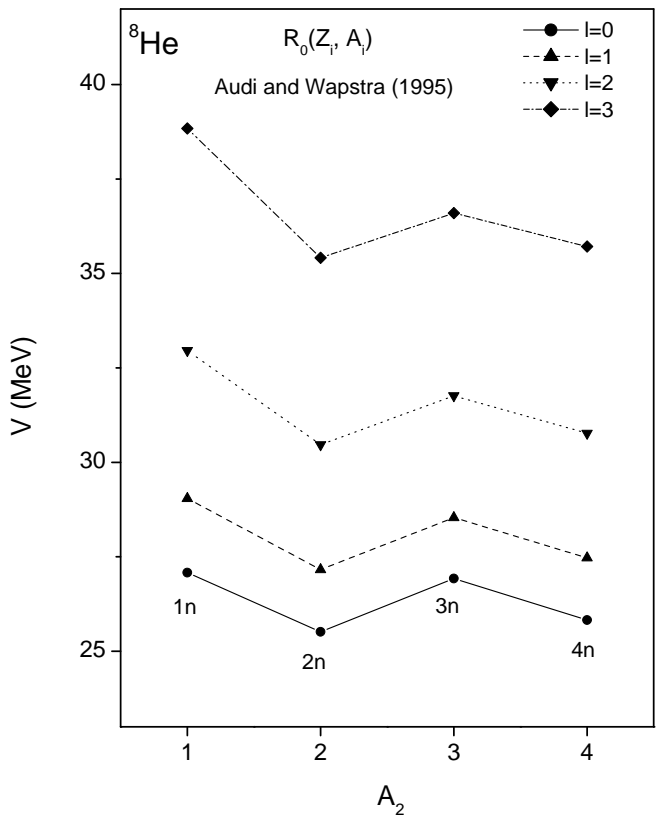
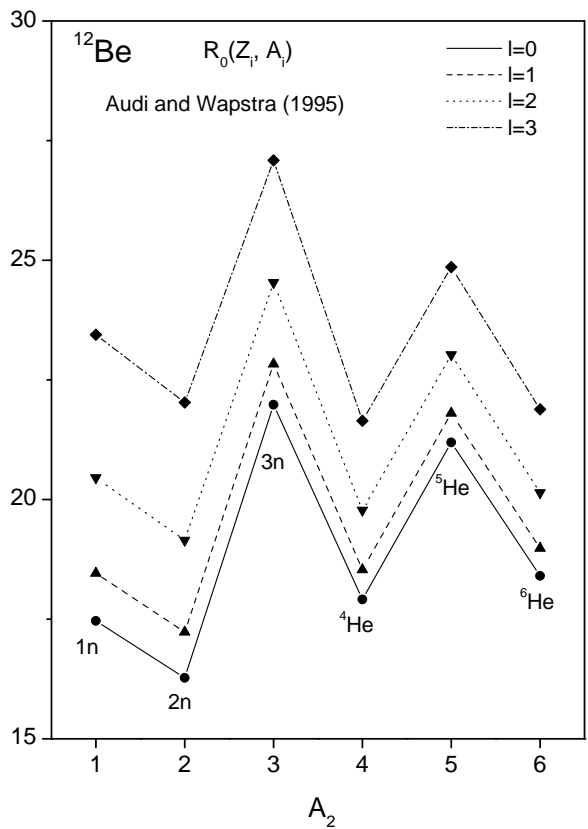
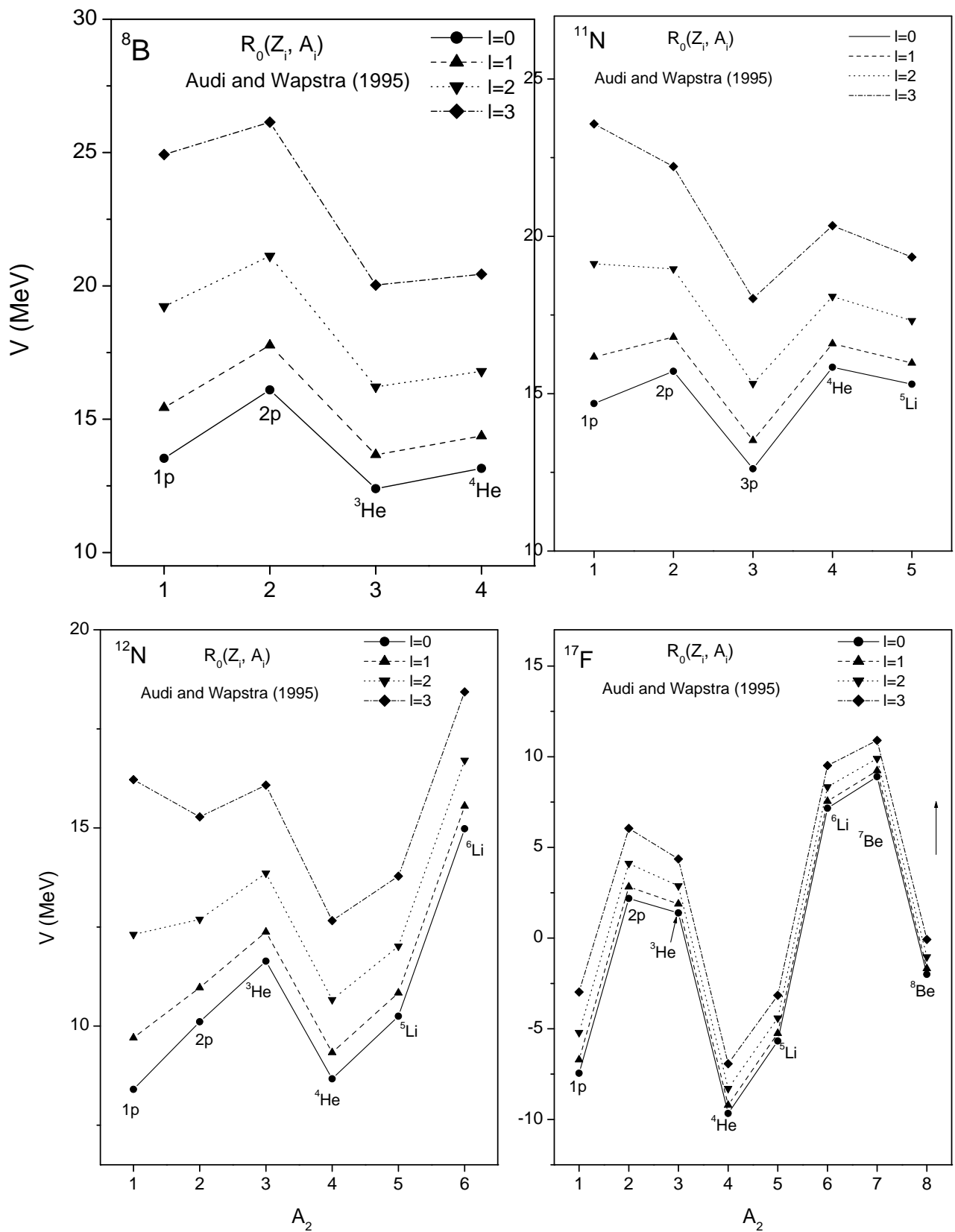
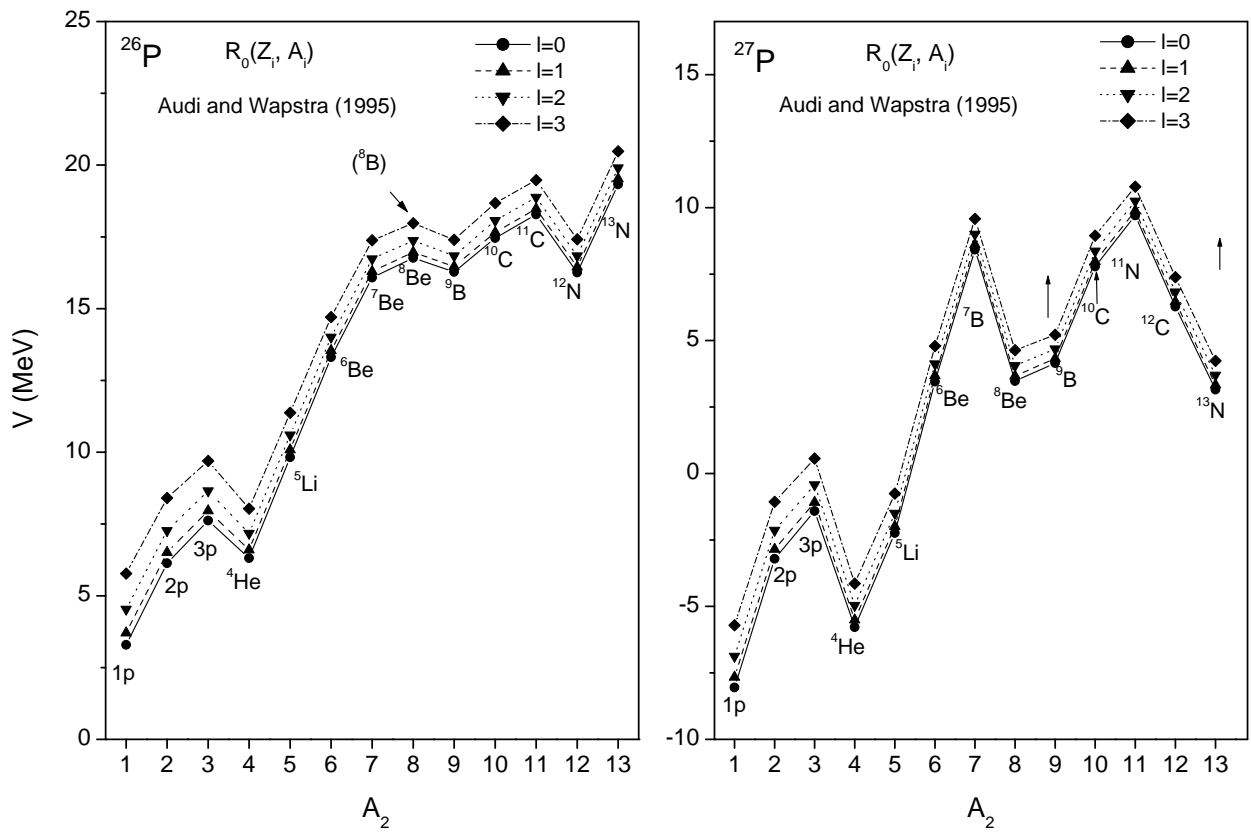
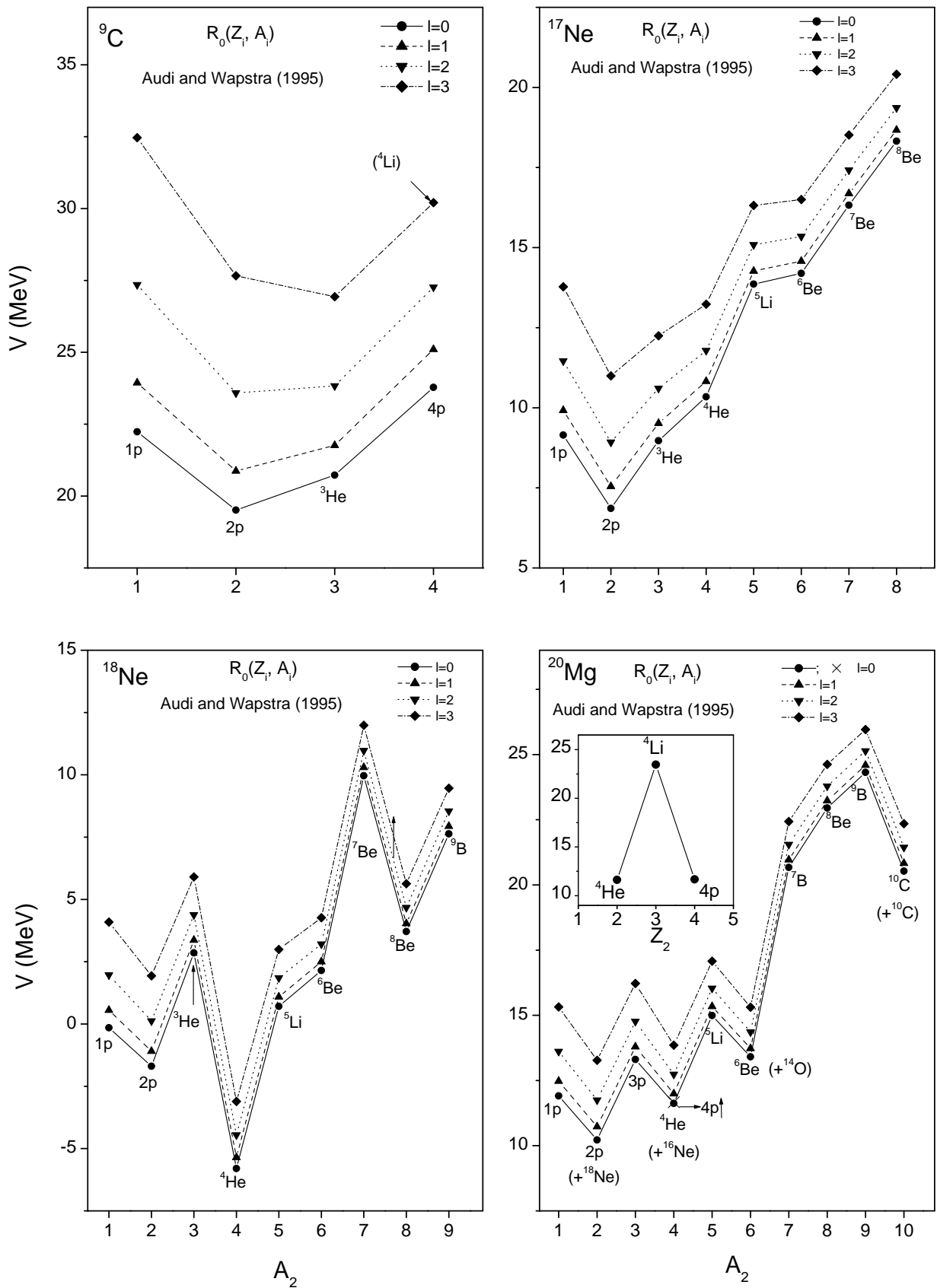
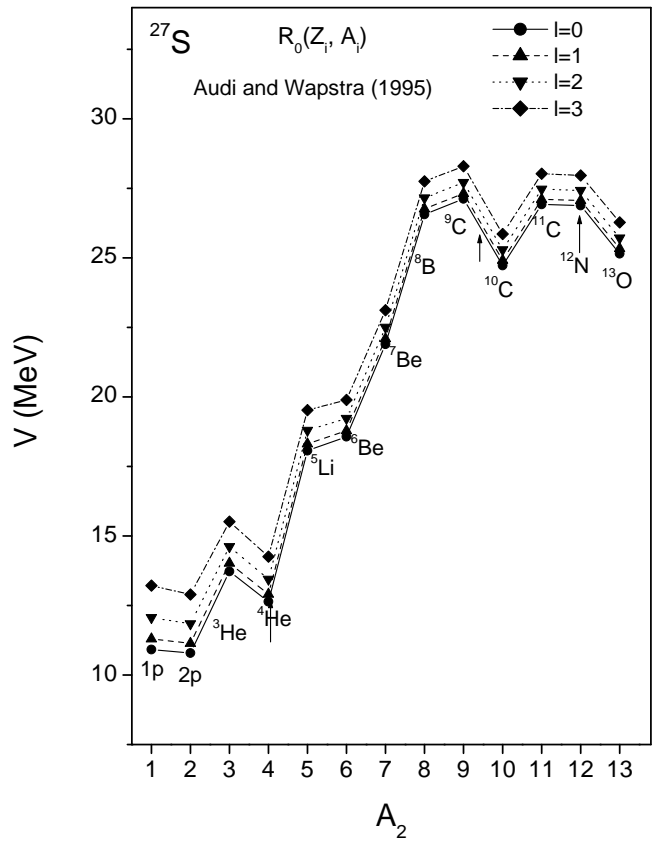
\end{document}